\documentstyle[12pt, preprint,aps,epsf]{revtex}

\def\m#1{$#1$}
\def\p{p}

\newcommand{\beq}{\begin{equation}}
\newcommand{\eeq}{\end{equation}}

\newcommand{\beqs}{\begin{eqnarray}}
\newcommand{\eeqs}{\end{eqnarray}}
\newcommand{\eqn}[1]{~(\ref{#1})}

\newcommand{\half}{\frac{1}{2}}

\newcommand{\grad}{\nabla}
\begin{document}
\title{A Condensation of Interacting Bosons in Two Dimensional Space}

\author{S. G. Rajeev
\\ Department of Physics and Astronomy\\
University of Rochester\\
Rochester,NY 14627 }

\maketitle

\abstract{			
We develop a theory of non-relativistic
 bosons  in two spatial dimensions  with a weak short range 
attractive interaction. In the limit as the range of the interaction
 becomes small, there is an ultra-violet divergence in the problem. We
 devise a scheme to remove this  divergence and  produce a completely
 finite formulation of the theory. This involves reformulating the
 dynamics in terms of a new operator whose eigenvalues give the 
{\it  logarithm} of the energy levels. Then, a  mean field theory is
 developed which allows us to describe the limit of a large number of
 bosons. The ground state  is a new kind of 
condensate (soliton) 
of bosons that breaks translation invariance spontaneously.
 The  ground state energy is negative and its magnitude 
grows {\it exponentially} with the number of
particles, rather than like a power law as for conventional many body systems.}

\begin{flushleft}
\end{flushleft}
\pagebreak

There is a well-known theory of Bose gases in one dimensional space with a 
delta-function interaction \cite{yang,huangbook}. However the two-dimensional 
bose system with {\it attractive} delta-function interactions 
 is logarithmically divergent even in the case of two bodies
\cite{jackiw,adhikariprl,2body,hendersonrajeev}. The hamiltonian of  the
theory is 
scale invariant so that the ground state energy is either zero or negative 
infinity; in the case of attractive interactions it is negative
infinity.  The two-body problem 
 has been intensely studied  even beyond perturbation theory
 and it is now understood how
 this divergence can be  
removed by a  renormalization. Recently we extended this analysis to the case
 of  the three
 body problem; while not exactly solvable, the problem can be 
reformulated in a finite form without using perturbation theory
\cite{hendersonrajeev}.

In this paper (for a more detailed discussion see Ref. \cite{trqm}) 
we will describe a non-perturbative, fully renormalized
 formalism for   the corresponding non-relativistic field 
theory; i.e., the case of an arbitrary number of bosons. We will then develop 
a mean field theory to obtain the ground state of the system. The results are 
striking: the ground state is a new kind of condensate (or soliton, a 
bound state of a large
 number of particles) of bosons, which breaks 
translation invariance spontaneously. Moreover, the binding energy grows
 exponentially with the number of particles. 

Of course, such a study of the 
binding of bosons is only possible because we now have a non-perturbative
renormalization method for this theory. This bosonic  field theory is 
asymptotically free (i.e., the running  coupling constant decreases  
logarithmically with energy) and hence provides a good testing  for  our new 
ideas of renormalization  \cite{trqm}. These ideas apply also 
to other asymptotically free theories such as the two-dimensional 
non-abelian Thirring model \cite{thirring} and perhaps,  even to four
dimensional gauge theories such as Quantum ChromoDynamics.

We begin with the naive hamiltonian of the system of bosons with an 
attractive self-interaction of zero range \cite{huangbook}:
\beqs
	H&=&\half\int |\nabla\hat\phi|^2 d^2x -
g\int \hat\phi^{\dag 2 }(x) \hat\phi^2(x)d^2x=\int {p^2\over
2}\phi^{\dag}(p) \phi(p){dp\over (2\pi)^2}\cr
& & -g
\int{dp_1dp_2dp_1'dp_2'\over (2\pi)^8}
		(2\pi)^2\delta(p_1+p_2-p_1'-p_2')
     \phi^{\dag}(p_1)\phi^{\dag}(p_2)\phi(p_1')\phi(p_2').\cr
\eeqs
Here, \m{\hat\phi(x)} is a complex scalar field in two dimensional space 
satisfying the usual canonical commutation relations; \m{\phi(p)=\int
e^{-ip\cdot x}\hat\phi(x)dx} is
its Fourier transform. In the units we are
 using, \m{\hat\phi} has dimensions of (length)$^{-1}$ and  \m{H} has
 dimensions 
of (length)$^{-2}$; it follows that the coupling constant \m{g} is
 dimensionless. This means that the lowest eigenvalue of \m{H} (in a
sector  with fixed number of particles ) is either zero or
infinite. If \m{g>0}, corresponding to an attractive potential, it is
known that the ground state energy is \m{-\infty} even in the case of
two particles \cite{2body}.

Following the standard ideas of field theory,
 we must regularize the problem by, for example, 
introducing a cut-off in the momentum; also,  the coupling
constant must be made a function of the cut-off in such a way that the
 ground state energy is finite. This is not always possible; for
 example the analogous problem in three dimensions requires an
 infinite number of coupling constants to be adjusted this way. When
 it is possible to remove all the divergences, by making a finite
 number of coupling constants depend on the cut-off, we say that the
 theory is renormalizable. We will show that our  bosonic field theory
 in two dimensions is renormalizable. Moreover, we will produce a
 completely finite reformulation of the theory {\it without the use of
 perturbation methods}. This allows us to ask questions about the bound
 states of the system, even the bound state of a large number of
 particles. a more detailed account of the methods used is in Ref. \cite{trqm}.

The regularized hamiltonian is, in momentum space,
\beq
	H_\Lambda=H_0+H_{1\Lambda}
\eeq
where
\beq
H_0=\int {p^2\over 2}\tilde\phi^{\dag}(p)\tilde \phi(p){dp\over (2\pi)^2}
\eeq
and
\beqs
	H_{1\Lambda}&=&-g(\Lambda)\int{dp_1dp_2dp_1'dp_2'\over (2\pi)^8}
\rho_{\Lambda}(p_1-p_2)\rho_\Lambda(p_1'-p_2')\cr
	& & 		(2\pi)^2\delta(p_1+p_2-p_1'-p_2')
     \phi^{\dag}(p_1)\phi^{\dag}(p_2)\phi(p_1')\phi(p_2').
\eeqs
Here \m{\rho_\Lambda} is a cut-off function that is equal to 
one near  the origin in  momentum space and vanishes rapidly at infinity.
 For example we could choose
\beq
	\rho_\Lambda(p)=\theta(|p|<\Lambda).
\eeq
We will see that this cut-off in the relative momentum is sufficient
to regularize the theory.

The solution of a quantum mechanical system is equivalent to
determining the resolvent (Green's function) of
its hamiltonian. Thus we would have an exact renormalization of our
theory if we can construct the resolvent \m{(H_\Lambda-E)^{-1}} in the
limit as \m{\Lambda\to \infty}. Our key result is that such a formula
for the resolvent can be obtained,
\beqs
	\lim_{\Lambda\to \infty}{1\over H_\Lambda-E}&:=&R(E)\cr
 &=&{1\over H_0-E}+
 {1\over 2}{1\over H_0-E}b^{\dag}
\left[{1\over 2\pi}\log{-E\over \mu^2}+W\right]^{-1} b {1\over
 H_0-E}.\label{principal}
\eeqs 
We will give below explicit formulas for the operators \m{b} and \m{W}. This
gives an indirect construction of the hamiltonian of the renormalized
theory as  the operator for which \m{R(E)=[H-E]^{-1}}. We
cannot get an explicit formula for the hamiltonian itself; instead we
express its resolvent in terms of the resolvent of an explicitly known
operator, \m{W}. The
dynamical information about the system is thus encoded in the integral
operator \m{W} (`the principal operator'); it 
 plays as important a role in our theory as the
hamiltonian does in usual physical theories.  
Once we reformulate the
theory this way, the standard approximation methods of quantum theory
can be applied to the principal operator. Later on we will in fact
apply mean field methods to get the ground state energy of the many
body system.

The operators \m{H_0}, \m{b}, \m{W} involve no dimensional parameters;
the only parameter in the theory is \m{\mu}. For example the bound
state energy levels of the system are given by the eigenvalues \m{w_a} 
of \m{W}:
\beq
	E_a=-\mu^2 e^{-2\pi w_a}.
\eeq
Once we determine one eigenvalue, it fixes \m{\mu} and then all other
eigenvalues are determined. For example, \m{-\mu^2} is the ground state
energy of the two-body system; once we know that the ground state of
the \m{n}-body system for any \m{n} is determined. The original
dimensionless coupling constant has been traded for the dimensional
constant \m{\mu} in the process of renormalization: this is
dimensional transmutation\cite{coleman}.

In order to get the above formula for the resolvent, we have to
overcome some technical difficulties.
The coupling constant appears in \m{H_\Lambda} 
 multiplicatively, which  makes it
difficult to perform a renormalization except, of course, in
perturbation theory. We will introduce some auxilliary  variables
 that will help us to rewrite the the theory in such
a way that coupling constant appears additively rather
multiplicatively. This is key to our method for exact renormalization.
Define creation-annihilation operators 
\m{\tilde\chi(p),\tilde\chi^\dag(p)} satisfying the product 
({\it not commutation} relations)
\beq
	\tilde\chi(\p)\tilde\chi^{\dag}(p')=(2\pi)^2\delta(\p-p'),\quad
\tilde\chi(\p)\tilde\chi(p')=0=\tilde\chi^{\dag}(\p)\tilde\chi^{\dag}(p').
\eeq
In addition, these operators commute with the bosonic fields. 
These relations are the \m{q\to 0} limit of the \m{q}-oscillators
studied by many authors (e.g., \cite{qosc}).
These relations have an obvious representation on \m{C\oplus L^2(R^2)}:
there is an empty state \m{|0>} and  states containing one auxilliary
particle  such
as  \m{\int v(p)\tilde\chi^\dag(p){dp\over (2\pi)^2}|0>}. Since the
product of the creation operators vanish, there can no more than one
such particle in {\it any} state: far more restrictive than the Pauli
exclusion principle for fermions. We dont have to  attach a
physical significance to these new auxilliary particles: they can be
viewed as a mathematical device that simplifies the analysis.
\footnote{ However there are examples (quarks or  ghosts) 
of  such auxilliary variables that  are important physically as well. We
propose that the auxilliary particles created by our operators 
\m{\chi,\chi^\dag} be called angels.}

 On the combined Hilbert space \m{{\cal
B}'} of bosons and the auxilliary particles 
 (\m{{\cal B}'={\cal B}\oplus {\cal B}\otimes
L^2(R^2)}, where \m{\cal B}) is the usual bosonic Fock space), we
define the operator:
\beqs
	 H_\Lambda'&=&H_0\Pi_0+
 \int
[dp_1 dp_2 dp_3]\rho_\Lambda(p_1-p_2)\cr
& & \phi^{\dag}(p_1)\phi^{\dag}(p_2)\chi(p_3 )(2\pi)^d\delta(p_1+p_2-p_3)+h.c.\bigg]+{1\over g(\Lambda)}\Pi_1.
\eeqs
Here,\m{\Pi_0,\Pi_1} are  the projection operators to the subspaces containing
zero or one  auxilliary particle    respectively:
\beq
	\Pi_0=\int [d\p]\chi(\p)\chi^{\dag}(\p),\quad
	\Pi_1=\int [d\p]\chi^{\dag}(\p)\chi(\p).
\eeq
We are interested in getting a formula for the resolvent of the
hamiltonian \m{H_\Lambda}. We will find it convenient to think in
terms of the auxilliary hamiltonian \m{H_\Lambda'} and 
 and the related operator  \m{R_\Lambda'=(H_\Lambda'-E\Pi_0)^{-1}}.

Let us   split the operators into \m{2\times 2} blocks according to
 auxilliary particle   number:
\beq
	 H_\Lambda'-E\Pi_0=\pmatrix{a&b^{\dag}\cr b&d\cr},\quad
	 R_{\Lambda}'(E)=
{1\over {\tilde H}_\Lambda-E \Pi_0}=\pmatrix{\alpha&\beta^{\dag}\cr
	      \beta&\delta}.
\eeq
with
\beq
	a,\alpha:{\cal B}\to {\cal B}, \quad b^{\dag},
\beta^{\dag}:{\cal B}\otimes
L^2(R^d)\to {\cal B}, d,\delta:{\cal B}\otimes L^2(R^d)\to {\cal B}\otimes
L^2(R^d).
\eeq
Two different ways of writing the elements of the inverse matrix will
 be useful (the proof is elementary):
\beq
	\left[a-b^\dag d^{-1}b\right]^{-1}=
\alpha=a^{-1}+a^{-1}b^{\dag}[d-ba^{-1}b^{\dag}]^{-1}b
 a^{-1},
\eeq
\beq
	\left[d-b  a^{-1}b^\dag\right]^{-1}=
\delta=d^{-1}+d^{-1}b[a-b^{\dag}d^{-1}b]^{-1}b^{\dag}  d^{-1},
\eeq
\beq
	-\delta b a^{-1}=\beta=-d^{-1}b\alpha.
\eeq
Using the left hand side of the formula for \m{\alpha} we can see that 
\beq
	\alpha=(H_\Lambda -E)^{-1}
\eeq
is the just the resolvent of the original bosonic system. The other
 way to write this quantity gives us
\beq
	{1\over H_\Lambda-E}={1\over H_0-E}+ {1\over
           2}{1\over H_0-E}b^{\dag}\Phi_\Lambda(E)^{-1} b {1\over
 H_0-E}.
\eeq
Here
\beqs
\Phi_\Lambda(E)&=&{1\over g(\Lambda)}-
\int
{dp_1dp_2dp_1'dp_2'\over (2\pi)^8}\rho_\Lambda(p_1-p_2)\rho_\Lambda(p_1'-p_2')\cr
& & 
\chi^{\dag}(p_1+p_2)\bigg[
  \phi(p_1)\phi(p_2){1\over H_0-E}
  \phi^{\dag}(p_1')\phi^{\dag}(p_2')\bigg]\chi(p_1'+p_2').
\eeqs
and
\beq
b^\dag=\int
[dp_1dp_2]\rho_\Lambda(p_1-p_2)\phi^{\dag}(p_1)\phi^{\dag}(p_2)
\chi(p_1+p_2).
\eeq

The point of this formula is that it expresses the resolvent of the
hamiltonian in terms of that of the free theory and the operator 
\m{\Phi_\Lambda(E)}. Moreover, the coupling constant appears additively
in \m{\Phi_\Lambda}; we will be able to separate the divergence in
\m{\Phi_\Lambda} and remove it by choosing \m{g(\Lambda)} to be an
appropriate function of \m{\Lambda}. Then the limiting operator
\m{\Phi(E)=\lim_{\Lambda\to \infty}\Phi_\Lambda(E)} will exist. The
roots of the equation \m{\Phi(E)|s>=0} will give the energy
levels. Thus we will obtain a completely finite reformulation of the
problem: what we have called elsewhere a transfinite formulation. It
is important that no where in this argument  we will  use perturbation theory.

In more detail, we can use the canonical commutation relations to
normal order the operators in \m{\Phi_\Lambda}. This will give
\beqs
   \Phi_\Lambda(E)&=&g^{-1}(\Lambda)-
\int
[dp_1dp_2dp_1'dp_2']\rho_\Lambda(p_1-p_2)\rho_\Lambda(p_1'-p_2')\cr
& & 
\chi^{\dag}(p_1+p_2)\bigg[\cr
& & \phi^{\dag}(p_1')\phi^{\dag}(p_2')
{1\over H_0+\omega(p_1')+\omega(p_2')+\omega(p_1)+\omega(p_2)-E}
		\phi(p_1)\phi(p_2)\cr
& & +4(2\pi)^d\delta(p_1-p_1')\phi^{\dag}(p_2'){1\over
H_0+\omega(p_1')+\omega(p_2')+\omega(p_2)-E}\phi(p_2)\cr
& & +2(2\pi)^d\delta(p_1-p_1')(2\pi)^d\delta(p_2-p_2')
{1\over H_0+\omega(p_1')+\omega(p_2')-E}\bigg]\cr
& & \chi(p_1'+p_2').
\eeqs
The only divergent term is the first one the square brackets. The 
divergence is removed if we choose
\beq
	g^{-1}(\Lambda)=\int \rho^2_\Lambda(\p){1\over \p^2+\mu^2}.
\eeq
Here we are trading the dimensionless coupling constant  \m{g} for the
dimensional constant \m{\mu}. The renormalized theory will depend on
\m{\mu} alone and not the cut-off \m{\Lambda} or on the coupling
constant: dimensional transmutation.

The operator \m{\Phi_\Lambda(E)} has a limit as \m{\Lambda\to \infty}:
\beqs
   \Phi(E)&=&\int [d\p]\chi^{\dag}(\p)\xi({\mu^2\over 2},H_0+{\p^2\over
2}-E)\chi(\p)\cr
& & -\int
[dp_1dp_2dp_1'dp_2']
\chi^{\dag}(p_1+p_2)\bigg[\cr
& & \phi^{\dag}(p_1')\phi^{\dag}(p_2')
 {1\over H_0+\omega(p_1')+\omega(p_2')+\omega(p_1)+\omega(p_2)-E}
		\phi(p_1)\phi(p_2)\cr
& & +4(2\pi)^d\delta(p_1-p_1')\phi^{\dag}(p_2'){1\over
H_0+\omega(p_1')+\omega(p_2')+\omega(p_2)-E}\phi(p_2)
\bigg]\cr
& & \chi(p_1'+p_2').
\eeqs
Here,
\beqs
	\xi({\mu^2\over 2},{\nu^2\over 2})&:=&\int[ d\p]
\bigg[{1\over p^2+\mu^2}-{1\over p^2+\nu^2}\bigg]\cr
&=& {1\over 4\pi}\ln{\nu^2\over \mu^2}
\eeqs
is a convergent integral.
We can rescale all the momemta by \m{\surd |E|} to get \footnote{ We
assume that \m{E<0} so that it is not in the spectrum of \m{H_0}. The
resolvent is defined elsewhere  by analytic continuation.}
\beq
	\Phi(\mu,E)=\bigg[{1\over 2\pi}\ln{-E\over \mu^2}+\;W\bigg].
\eeq
(From this point on our  momentum variables \m{p,p'} and position
variables \m{x} are {\it dimensionless}.) The principal operator \m{W} 
is,
 (\m{\omega(p)={p^2\over 2}})
\beqs
	W&=&{1\over 2\pi}\int [d\p]\chi^{\dag}(\p)
\log\bigg[H_0+\omega(p)+1\bigg]\chi(\p)\cr
& & -\int
[dp_1dp_2dp_1'dp_2']
\chi^{\dag}(p_1+p_2)\bigg[\cr
& & \phi^{\dag}(p_1')\phi^{\dag}(p_2')
 {1\over H_0+\omega(p_1')+\omega(p_2')+\omega(p_1)+\omega(p_2)+1}
		\phi(p_1)\phi(p_2)\cr
& & +4(2\pi)^d\delta(p_1-p_1')\phi^{\dag}(p_2'){1\over
H_0+\omega(p_1')+\omega(p_2')+\omega(p_2)+1}\phi(p_2)
\bigg]\cr
& & \chi(p_1'+p_2').
\eeqs 
This proves the formula \eqn{principal} for the resolvent.

Thus the eigenvalues of the operator \m{W} will determine
the energy levels. In fact,
\beq
	W|\psi>=w|\psi>,\quad  E=-{\mu^2}e^{-2\pi w}.
\eeq
The problem is completely
well-posed in terms of the operator \m{W}; it is a replacement for the
hamiltonian of the theory. In fact there are many subtleties in the
definition of the hamiltonian itself (e.g., domain self-adjointness), 
all of which can be avoided by
working with \m{W}. We call this remarkable new operator that embodies
the dynamics of the system the `Principal Operator'. Roughly speaking
it represents  the logarithm of energy.

Now that we have a finite formulation of the theory, we can look for
analogues  of  all the approximation methods of quantum theory. In
particular, we can look for a mean field theory. The idea is to use a
variational principle to estimate the lowest eigenvalue of \m{W} in
the sector with \m{n} bosons. Now
\m{W} acts in the Hilbert space of \m{n} bosons  and one auxilliary
particle. The
variational ansatz corresponding to mean field theory 
 is the one where all the bosons are in the state \m{u} and the angel
is in the state \m{v}:
\beq
	|u,v>=\int {dp_1\cdots dp_n dp\over
(2\pi)^{2n+2}}u(p_1)\cdots u(p_n)v(p)\phi^\dag(p_1)\cdots
\phi^\dag(p_n)\chi^\dag(p)|0>.
\eeq
The expectation value of \m{W} in this state becomes (for large \m{n})
the `principal function'
\beqs 	
 U&=&{1\over 2\pi}\int [dp] |v(p)|^2\log[n h_0(u)+\omega(p)+1]\cr
  & & -\int[dp_1dp_2dp_1'dp_2']v(p_1+p_2)^*v(p_1'+p_2')\cr
  & & \bigg[{u^*(p_1')u^*(p_2')u(p_1)u(p_2)n(n-1)
         \over
     nh_0(u)+\omega(p_1')+\omega(p_2')+\omega(p_1)+\omega(p_2)+1}\cr
 & & +4{(2\pi)^2\delta(p_1-p_1')nu^*(p_2')u(p_2)
      \over nh_0(u)+\omega(p_1')+\omega(p_2')+\omega(p_2)+1}\bigg]
\eeqs
where 
\beq
	h_0(u)=\int |u(p)|^2\omega(p)[dp].
\eeq
This \m{U} is to be minimized subject to the normalization conditions
\beq
	\int |u(p)|^2[dp]=\int |v(p)|^2[dp]=1.
\eeq 
If we keep just the leading terms as \m{n\to\infty}, 
\beq
	U=-n{\bigg|\int
v^*(p_1+p_2)u(p_1)u(p_2)[dp_1dp_2]\bigg|^2\over h_0(u)}+{1\over
2\pi}\log n-{1\over 2\pi}\log f_1(u)+{\rm O}({ n^{-1}}).
\eeq
The functional  \m{f_1(u)} (which is independent of \m{n}) 
can also determined as above, but it takes a bit
more  work. Thus we have the  approximate 
variatonal principle for
the ground state energy of a system of \m{n} bosons:
\beq
	E_n=-\mu^2{e^{2\pi n \over \xi}\over n}\big[C_1+O({1\over
n})\big]
\eeq
where
\beq
	\xi=\inf_{u,v}{
\int|v(p)|^2[dp]\int|u(p)|^2[dp]\int\omega(p)|u(p)|^2[dp]
\over\bigg|\int
v^*(p_1+p_2)u(p_1)u(p_2)[dp_1dp_2]\bigg|^2 }.
\eeq
Also, \m{C_1} is a constant; it is   not necessary to determine it for
the leading behavior in the large \m{n} limit of energy.
We can eliminate \m{v} rather easily to get an equivalent form
(written in terms of the position space wavefunction 
\m{\tilde u(x)=\int u(p)e^{ipx}{dp\over (2\pi)^2})}):
\beq
	\xi=
\inf_{\tilde u}I[u]
\eeq
where,
\beq
I[\tilde u]={\int|\tilde u(x)|^2dx\int\half|\nabla \tilde u(x)|^2dx
\over \int
\tilde|\tilde u(x)|^4 dx }.
\eeq 
All the variables in our problem are dimensionless. The minimization
of \m{\log I[u]} gives the partial differential equation
\beq
\nabla^2\tilde u -\beta \tilde u +\lambda |\tilde u|^2 \tilde u=0
\eeq
where,
\beq
	\beta={\int |\grad \tilde u|^2 d^2x \over \int |\tilde u|^2 d^2x},\quad 
\lambda={\int |\grad \tilde u|^2 d^2x \over \int |\tilde u|^4 d^2x}.
\eeq
This nonlinear Schrodinger equation has been shown to have solutions  in
\cite{berger},Theorem 6.7.25.( Because of the scale covariance it is
enough to find solutions with one positive value of \m{\beta} and
\m{\lambda}.) We expect the ground state to have at least circular
symmetry; we get a sort of nonlinear Bessel's equation:
\beq
	v''(r)+{1\over 4r^2}v(r)+{v^3(r)\over r}=v(r)
\eeq
with \m{\tilde u(x)=\surd r v(r)} and \m{r=|x|}. It is easy to solve
this equation numerically; we plot the result in the figure. The value
of \m{\xi} we get is about 12. The solution of course breaks
translation invariance spontaneously but not rotation invariance. It
is in fact a `soliton' in the sense of field theory. 

\centerline{\epsfbox{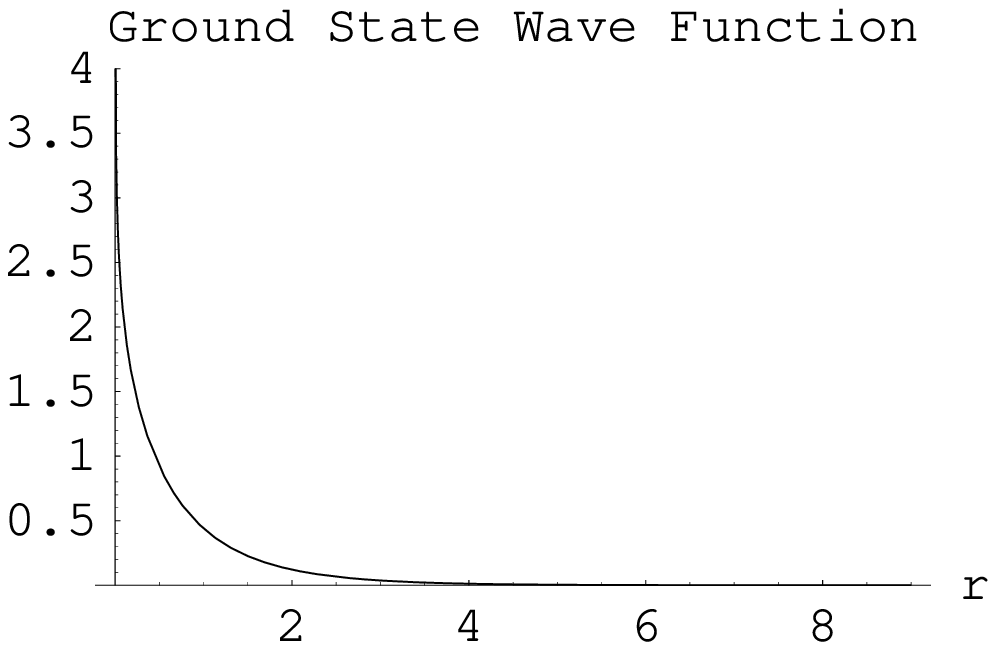}}

It follows that the ground state energy (binding energy) 
 grows exponentially with the
number of particles:
\beq
		E_n=-\mu^2e^{{\pi\over 6}n+\cdots}.
\eeq
Here, \m{-\mu^2} is the binding energy of the two boson system; the
 dots denote terms of lower order in the number of bosons. 
 At zero temperature, 
essentially all the avaliable particles will fall into this
 bound state. That is the sense in which this state is a condensate.

The wavefunction was determined above in terms of a dimensionless
position variable. By putting back the factor \m{\surd|E|} we scaled
out, we see that the size of the condensate behaves like 
\m{|E_n|^{-\half}\sim \mu^{-1}e^{-{\pi\over 12}n}}; it shrinks with
increasing number of particles. Although the results above are
asymptotic for large \m{n}, we expect (by analogy with mean field
theory in atomic physics) that  they will hold true even for moderate
values of \m{n } of about 3 or more. 

It may be possible to observe such a condensation in cold optical 
traps of dilute bose gases, if we can trap the atoms close to some
surface \cite{Raizen}. Even without a confining potential in the plane of the
surface, the atoms will form a `self-trapped' bound state. It may be
possible to tune the attractive force between atoms  using
Feshbach resonances \cite{Bigelow}. Then, even with an exponentially small size for
the condensate, the atoms can be  far apart and the approximation of
zero range will be valid. Whether large particle numbers in the
condensate can be achieved is not clear yet, however. There might also be a 
realization in terms of vortices in superconductors which are on the boundary 
between type I and type II \cite{trqm}.

In the case of the three dimensional attractive Bose gas, even after
the above renormalization the energy is not bounded below\cite{trqm};
our methods fail. In fact there are indications ( both theoretical and
experimental)  that in this case,
the system is truly unstable \cite{huang3d,hulet}.

\end{document}